\documentclass[useAMS,usenatbib]{mn2e}
\usepackage{epsfig,natbib}
\usepackage{amssymb}
\usepackage{graphicx}
\usepackage{placeins}
\usepackage{times}
\usepackage{epsfig}

\title[
Galaxy Counterparts of metal-rich DLAs
]
{
Galaxy Counterparts of metal-rich Damped Lyman-$\alpha$ Absorbers - I:
The case of the $z=2.35$ DLA towards Q\,2222$-$0946
\footnotemark[1]\thanks{Based on
observations carried out at the European Organisation for Astronomical Research
in the Southern Hemisphere, Chile, under ESO program 084.A-0303(A).}
}
\author[J. P. U. Fynbo et al.]{J. P. U. Fynbo$^{1}$\thanks{E-mail:
jfynbo@dark-cosmology.dk},
P. Laursen$^{1}$,
C. Ledoux$^{2}$,
P. M\o ller$^{3}$,
A. K. Durgapal$^{4}$,
P. Goldoni$^{5,6}$,\newauthor
B. Gullberg$^{1}$,
L. Kaper$^{7}$,
J. Maund$^{1}$,
P. Noterdaeme$^{8}$,
G. \"Ostlin$^{9,10}$,
M. L. Strandet$^{1}$,\newauthor
S. Toft$^{1}$,
P. M. Vreeswijk$^{1}$,
T. Zafar$^{1}$
\\
$^{1}$Dark Cosmology Centre, Niels Bohr Institute, Copenhagen University, Juliane Maries Vej 30, 2100 Copenhagen O, Denmark\\
$^{2}$European Southern Observatory, Alonso de C\'ordova 3107, Casilla 19001, Vitacura, Santiago 19, Chile\\
$^{3}$European Southern Observatory, Karl-Schwarzschildstrasse 2, D-85748 Garching, Germany\\
$^{4}$Department of Physics, DSB Campus, Kumaun University, Nainital, Uttarakhand, India\\
$^{5}$Laboratoire Astroparticule et Cosmologie, 10 rue A. Domon et L. Duquet, F-75205 Paris Cedex 13, France\\
$^{6}$Service d'Astrophysique, DSM/DAPNIA/SAp, CEA-Saclay, F-91191 Gif-sur-Yvette, France\\
$^{7}$Astronomical Institute "Anton Pannekoek", University of Amsterdam, Kruislaan 403, 1098 SJ Amsterdam, The Netherlands\\
$^{8}$Departamento de Astronom\'ia, Universidad de Chile, Casilla 36-D, Santiago, Chile\\
$^{9}$Department of Astronomy, Stockholm University, AlbaNova University Center, 10691 Stockholm, Sweden\\
$^{10}$Oscar Klein Centre for Cosmoparticle Physics, Department of Astronomy, Stockholm University, Stockholm, Sweden
}
\begin{document}

\date{Accepted . Received ; in original form}

\pagerange{} \pubyear{2010}

\maketitle

\label{firstpage}

\begin{abstract}
  We have initiated a survey using the newly commissioned X-shooter
  spectrograph to target candidate relatively metal-rich damped
  Lyman-$\alpha$ absorbers (DLAs). Our rationale is that
  high-metallicity DLAs due to the luminosity-metallicity relation
  likely will have the most luminous galaxy counterparts. In addition,
  the spectral coverage of X-shooter allows us to search for not only
  Lyman-$\alpha$ (Ly$\alpha$) emission, but also rest-frame optical
  emission lines. We have chosen DLAs where the strongest rest-frame
  optical lines ([OII], [OIII], H$\beta$ and H$\alpha$) fall in the
  NIR atmospheric transmission bands. In this first paper resulting
  from the survey, we report on the discovery of the galaxy
  counterpart of the $z_{\rm abs} = 2.354$ DLA towards the $z=2.926$
  quasar Q\,2222$-$0946. This DLA is amongst the most metal-rich $z>2$
  DLAs studied so far at comparable redshifts and there is evidence
  for substantial depletion of refractory elements onto dust grains.
  We measure metallicities from ZnII, SiII, NiII, MnII and FeII of
  $-0.46\pm0.07$, $-0.51\pm0.06$, $-0.85\pm0.06$, $-1.23\pm0.06$, and
  $-0.99\pm0.06$, respectively. The galaxy is detected in the
  Ly$\alpha$, [OIII] $\lambda$4959, $\lambda$5007 and H$\alpha$
  emission lines at an impact parameter of about 0.8 arcsec (6 kpc at
  $z_{\rm abs} = 2.354$). Based on the H$\alpha$ line, we infer a
  star-formation rate of 10 $M_{\sun}$ yr$^{-1}$, which is a lower
  limit due to the possibility of slit-loss. Compared to the recently
  determined H$\alpha$ luminosity function for $z=2.2$ galaxies the
  DLA-galaxy counterpart has a luminosity of  $L\sim0.1L^*_\mathrm{H\alpha}$. 
  The emission-line ratios are 4.0 (Ly$\alpha$/H$\alpha$) and 1.2
  ([OIII]/H$\alpha$). In particular, the Ly$\alpha$ line shows clear
  evidence for resonant scattering effects, namely an asymmetric,
  redshifted (relative to the systemic redshift) component and a much
  weaker blueshifted component. The fact that the blueshifted
  component is relatively weak indicates the presence of a galactic
  wind.

  The properties of the galaxy counterpart of this DLA is consistent
  with the
  prediction that metal-rich DLAs are associated with the most
  luminous of the DLA-galaxy counterparts. 
\end{abstract}


\section{Introduction}

A central aspect of the history of the Universe is the formation and
evolution of galaxies, and in particular their gradual build-up of
metallicity (Pei \& Fall 1995; Cen \& Ostriker 1999; Pettini 2006;
Sommer-Larsen \& Fynbo 2008). While the study of galaxies in the early
universe has seen great progress through the identification of large
samples of $z\gtrsim3$ Lyman-Break Galaxies (LBGs) and other
emission-selected galaxies (e.g., Steidel et al.\ 1996, 2004; Bouwens
et al.\ 2009), the study of the metals at the same cosmic epoch has
seen equally large progress but mostly based on the study of
$z\gtrsim2$ DLAs in the sightlines towards bright quasars (e.g., Wolfe
et al.\ 2005; Pontzen \& Pettini 2009). Thanks to the SDSS the sample
of high-redshift DLAs is now larger than 1000 (Prochaska et al.\ 2005;
Noterdaeme et al.\ 2009). In order to be able to form a complete
picture of galaxy evolution one must find a way to combine the
information inferred from high-redshift galaxies studied in emission
and absorption, and it is a perplexing fact that there is almost no
{\it observational} overlap between the two samples 
(e.g., Fynbo et al.\ 1999; M\o ller et al.\ 2002).

It is not obvious how one may combine the two samples because they
have very different selection biases. LBG samples are all
flux-limited, and therefore they carry information only about objects
at the bright end of the luminosity function. DLAs are selected based
on the probability that a random sightline passes through it, and they
are therefore selected with a weight proportional to the absorption
cross-section (area) they span on the sky. This area is known,
locally, to scale with the luminosity to a given power (the so-called
Holmberg and Bosma relations; Wolfe et al.\ 1986 and references
therein; see also Zwaan et al.\ 2005). On the reasonable assumption
that similar relations were in place at $z=3$, and combining this with
the slope of the faint end of the UV luminosity function, one finds that
most DLAs are selected from the faint end of the luminosity function
(Fynbo et al.\ 1999, 2008; Haehnelt et al.\ 2000; Schaye 2001; Pontzen
et al.\ 2008). Hence, the plausible connection between DLAs and LBGs
is that they are drawn from the same overall population of high-$z$
galaxies but that in the mean they are picked from two opposite ends
of the high-$z$ galaxy UV luminosity function.

To better establish the validity of this picture, more detections of
DLAs in emission are required. Detection of emission from DLAs has
been vigorously pursued for more than 20 years (e.g., Smith et al.\
1989), but with limited success. Since 1993 only two bona-fide
high-redshift DLA galaxies have been detected in emission (M\o ller et
al.\ 2002, 2004; Heinm\"uller et al.\ 2006). In addition, counterparts
of either $z_{\rm abs} \approx z_{\rm em}$ DLAs or sub-DLAs have been
published (M\o ller \& Warren 1993, 1998; Warren \& M\o ller 1996;
Djorgovski et al. 1996; Leibundgut \& Robertson 1999; Fynbo et al.\
1999; Adelberger et al.\ 2006). At least ten times more systems have
been searched for in emission with no detection (e.g., Charlot \& Fall
1991 and references therein; Lowenthal et al.\ 1995; Colbert \& Malkan
2002; Kulkarni et al.\ 2006). In addition to published non-detections
many non-detections have not been published, e.g., non-detections in
the survey conducted with FORS\,1 at the VLT 1999--2000 (PI M\o ller),
and the FLAMES/IFU survey also conducted at the VLT 2003--2005 (PIs
Leibundgut and Zwaan)). More recently, tentative detections based on
integral-field unit spectrographs have been reported (Christensen et
al.\ 2007) and one of these has been confirmed with long-slit
spectroscopy (Christensen, private communication). The low success
rate has been thought to be the result of attenuation of Ly$\alpha$
photons by dust grains (Charlot \& Fall 1991), but as discussed above
it is also plausibly explained in terms of the prediction that DLA
galaxies must in the mean be drawn from the faint end of the
luminosity function.

M\o ller et al.\ (2004) and Ledoux et al.\ (2006) provide evidence
that DLA galaxies obey luminosity-metallicity and velocity-metallicity
relations with similar slopes as in the local Universe. Moreover, the
DLA galaxies that have been detected in emission (both Ly$\alpha$,
broad band, and for those with redshifts in a range allowing
observations in NIR atmospheric windows, also in [OIII] emission) are
amongst the most metal-rich DLAs (M\o ller et al.\ 2004; Weatherley et
al.\ 2005). Fynbo et al.\ (2008) describe a model that reconciles the
known properties of DLAs, LBGs and also Gamma-Ray Burst host galaxies
based on scaling relations known from the local Universe (Luminosity
Function, Holmberg/Bosma relation, luminosity-metallicity relation and
metallicity gradients). There is no doubt that reality is more complex
than this simple picture (see, e.g., Zwaan et al.\ 2008 and Tescari et
al.\ 2009 for a discussion of the likely importance of galactic winds
for the kinematical properties of DLAs). Nevertheless, the basic
ingredients in this picture have been found also in hydrodynamical
simulations of DLAs at $z\approx3$ (Pontzen et al.\ 2008).

To test this picture, we have embarked on a survey targeting
high-metallicity DLAs in order to search for the galaxy counterparts.
The project is carried out using our guaranteed time on the newly
commissioned X-shooter spectrograph on the European Southern
Observatory (ESO) Very Large Telescope (VLT) on Cerro Paranal in
Chile. X-shooter is an echelle spectrograph with three arms covering
the full spectral range from the atmospheric cut-off around 2950 \AA \
to the K-band. From the SDSS DLA sample of Noterdaeme et al.\ (2009),
we have selected DLAs with a rest-frame equivalent width (EW) of the
SiII $\lambda1526$ line larger than 1 \AA. This is a good indication
that the metallicity is high, i.e. likely larger than 0.1 solar (e.g.,
Prochaska et al.\ 2008, their figure 6; Kaplan et al.\ 2010). 
Among these DLAs, we selected
the subsample with well-detected FeII\,$\lambda$2344, $\lambda$2374,
and $\lambda$2382 lines. For local starburst galaxies, the strength of
the Ly$\alpha$ emission line is controlled by a complex interplay
between geometry, kinematical properties of the gas and dust content
(see, e.g., Atek et al.\ 2009 and references therein). For this
reason, we only include DLAs with redshifts close to 2.4 which places
the H$\beta$, [OII], [OIII] and/or H$\alpha$ emission lines from the
intervening DLA in the NIR transmission windows. At these redshifts,
the spectral range covered by X-shooter extends from the Lyman-limit
to H$\alpha$ and we are not relying on Ly$\alpha$ alone for detecting
the galaxy counterpart. In this paper, we report on the first target
observed in the program, namely SDSS J\,222256.11$-$094636.2
(Q\,2222$-$0946 in the following), which has an intervening DLA at a
redshift of $z_{\rm abs} = 2.354$ (Noterdaeme et al.\ 2009).

Throughout this paper, we assume a flat cosmology with
$\Omega_{\Lambda}=0.70$, $\Omega_m = 0.30$, and a Hubble constant of
$H_0 = 70$ km s$^{-1}$ Mpc$^{-1}$.

\section{Strategy and Observations}

\begin{figure}
\includegraphics[width=0.48\textwidth]{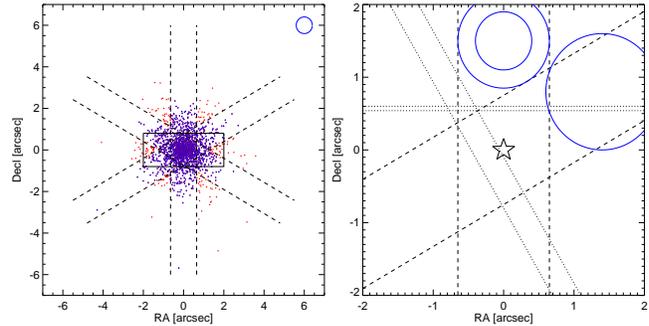}
\caption{{\it Left panel:} A simulation based on the model in Fynbo et
  al.\ (2008) on how many DLA-galaxy counterparts with metallicities
  above 0.1 solar will be missed by our three X-shooter slit
  positions. In this coordinate system, the background QSO is located
  in the origo. The three slits are marked with dashed lines and the
  centres of 5000 simulated DLA-galaxy counterparts are marked with
  blue (if inside one of the slits) or red (if they fall outside all
  three slits). Slightly above 90\% of the galaxy centres are covered
  by at least one slit. In the upper right corner, we show the size of
  the seeing disk for a seeing of 0.8 arcsec. Due to seeing,
  substantial flux will be detected even for some galaxies with
  centres outside of our slits. The full-drawn rectangle illustrates the
  field-of-view of the X-shooter IFU. {\it Right panel:} Here we show
  the triangulation of the DLA-galaxy counterpart towards
  Q\,2222$-$0946. The dashed lines again show the $PA=0^\mathrm{o}$
  and $PA=-60^\mathrm{o}$ slits. The position of the QSO is marked
  with a star and the seeing (in the B-band) for the three spectra 
  are illustrated
  with blue circles. The dotted lines mark the 1$\sigma$ regions for
  the measured impact parameter in the two slits. The overlapping
  region lies close to the edge of both slits (formally outside the
  $PA=-60^\mathrm{o}$ slit) at a position angle of 40$^\mathrm{o}$ and
  with an impact parameter of 0.8 arcsec. \label{fig:strategy} }
\end{figure}

Q\,2222$-$0946 was observed on October 21 2009 with X-shooter. Our
strategy was to observe the QSO at the position angles (PAs)
60$^\mathrm{o}$, $-60^\mathrm{o}$, and 0$^\mathrm{o}$ (all East of
North), 1 hr each, using a 1.3 arcsec slit in the UVB arm and 1.2
arcsec slits in the VIS and NIR arms. The resulting resolving powers
are 4700 (UVB), 6700 (VIS), and 4400 (NIR) respectively (as measured
from sky lines in the spectra). With this strategy, we cover the field
of view close to the QSO and will be able to determine the impact
parameter and position angle of galaxy counterparts using
triangulation (see M\o ller et al.\ 2004 for an example of such a
triangulation). Based on the model by Fynbo et al.\ (2008) mentioned
above, we predict that less than 10\% of the galaxy counterparts will
fall outside all three slits (see Fig.~\ref{fig:strategy}). With a
full-drawn rectangle, we illustrate the field covered by the X-shooter
IFU. With a single IFU pointing centred on the QSO, 30\% of the
DLA-galaxy centres will be lost. We prefer to use long slits as this
strategy provides superior data for the QSO itself and the data
analysis is more robust.

Due to an error in the execution of our Observing Blocks at the
telescope, the object was observed twice at PA$=0^\mathrm{o}$ and once
at PA$=-60^\mathrm{o}$. Hence, we did not get a spectrum at
PA$=60^\mathrm{o}$. The observing conditions were variable with a
seeing degrading from 0.8 to 1.6 arcsec (as measured in the B-band).
The airmass ranged from 1.0 to 1.2 and there were no clouds.

We processed the spectra using a preliminary version of the X-shooter
data reduction pipeline (Goldoni et al.\ 2006). The pipeline performs
the following actions. First, the raw frames are corrected for the
bias level (UVB and VIS) and dark current (NIR). Then, after
background subtraction, cosmic ray hits are detected and corrected
using the method developed by van Dokkum (2001) and sky emission lines
are subtracted using the Kelson (2003) method. After division by the
flat field, the orders are extracted and rectified in wavelength space
using a wavelength solution previously obtained from calibration
frames. The orders are then merged and in the overlapping regions the
merging is weighted by the errors which are being propagated in the
process. From the resulting 2D merged spectrum, a one-dimensional
spectrum is extracted at the source's position. The one-dimensional
spectrum with the corresponding error file and bad-pixel map is the
final product of the reduction. Intermediate products such as the sky
spectrum and individual echelle orders (with errors and bad-pixel
maps) are also produced.

To perform flux calibration, we extracted with the same procedures a
spectrum of the flux standard BD +17$^{\rm o}$ 4708 (Bohlin \&
Gilliland 2004). This spectrum was divided by the flux table of the
same star from the CALSPEC HST database (Bohlin
2007\footnote{http://www.stsci.edu/hst/observatory/cdbs/calspec.html})
to produce the response function. The response was then interpolated
where needed in the atmospheric absorption bands in the VIS and NIR
spectra and applied to the spectrum of the source. No telluric
correction was applied. We compare our flux calibration to the flux
calibrated spectrum from the Sloan survey (Adelman-McCarthy et al.\ 2009) and find that our calibration gives 30\% larger fluxes compared
to the Sloan calibration, but the shapes of the two spectra match very
well. We have chosen to rescale our spectrum so that is matches the
Sloan spectrum.

\section{Results}

\subsection{Emission properties of the DLA-galaxy counterpart}

\begin{figure}
\includegraphics[width=0.48\textwidth]{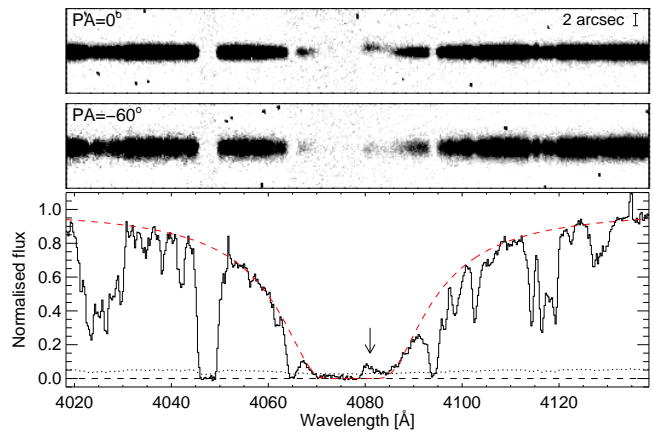}
\caption{The damped Lyman-$\alpha$ absorption line in one and two
  dimensions. The upper two plots show the two-dimensional spectra
  from the $PA=0^\mathrm{o}$ spectrum taken under the best seeing and
  the $PA=-60^\mathrm{o}$ spectrum. The Ly$\alpha$ emission from the
  DLA-galaxy counterpart is seen close to the red wing of the DLA line
  offset by 0.57 arcsec above the QSO trace in the $PA=0^\mathrm{o}$
  spectrum and 0.20 arcsec below the trace in the $PA=-60^\mathrm{o}$
  spectrum. In the lower plot, we show the one-dimensional spectrum
  combining all three spectra. 
  The noise
  spectrum is plotted as a dotted line. The dashed line shows the
  result of a Voigt-profile fit to the DLA line from which a column
  density of $\log N=20.65\pm0.05$ is inferred. The Ly$\alpha$
  emission in the trough is indicated with an arrow. Note that in this
  one-dimensional spectrum the Ly$\alpha$ emission is not entirely
  recovered as the spectrum is extracted from the combined QSO spectrum
  along the QSO trace and the
  Ly$\alpha$ emission is spatially offset from the QSO trace.
  \label{fig:dla} }
\end{figure}

\begin{figure}
\includegraphics[width=0.48\textwidth]{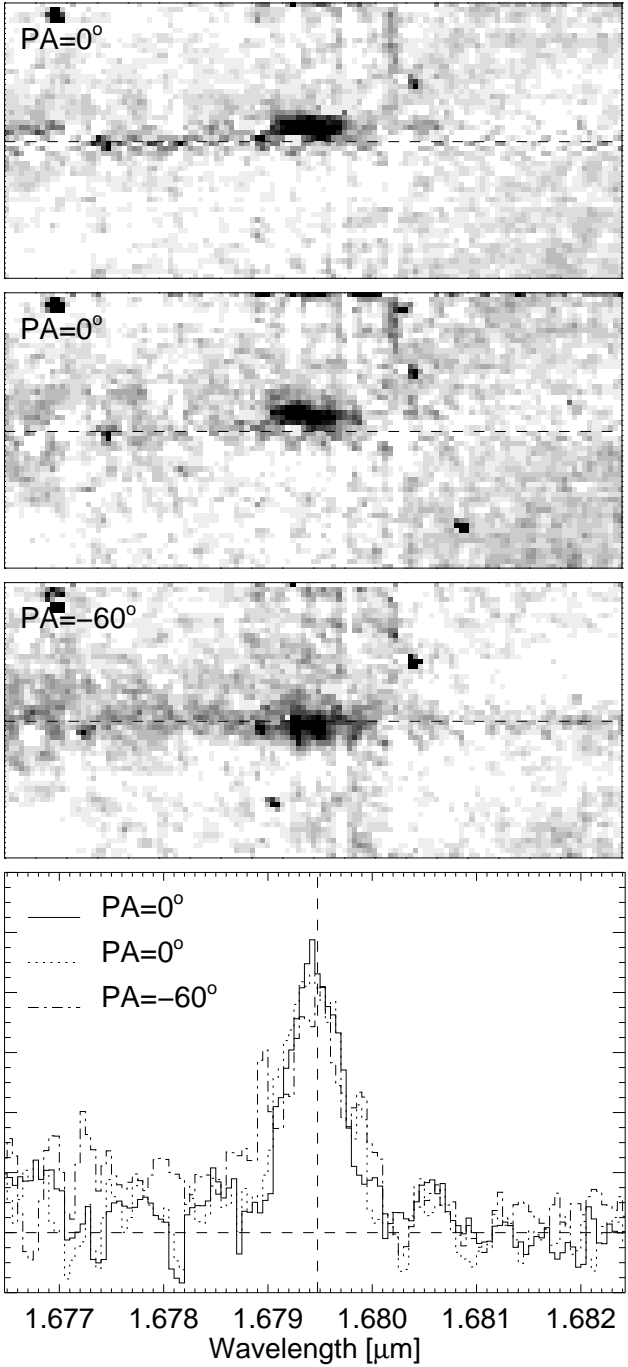}
\caption{The [OIII] $\lambda$5007 emission line from the DLA-galaxy
  counterpart. The upper three figures show the two-dimensional spectra
  after SPSF subtraction of the QSO continuum. The position of the QSO
  trace (which has been subtracted) is marked by a horizontal dashed
  line. In the bottom panel, we plot the one-dimensional spectrum from
  each of the above three spectra with dotted, full-drawn and
  dot-dashed lines. The vertical dashed line shows the predicted
  position of the [OIII] line from the redshift determined from
  low-ionisation absorption lines. The horizontal dashed line
  indicates the zero-flux level. The spectra have been rescaled to
  have the same peak count-level. \label{fig:oiii} }
\end{figure}

\begin{figure}
\includegraphics[width=0.48\textwidth]{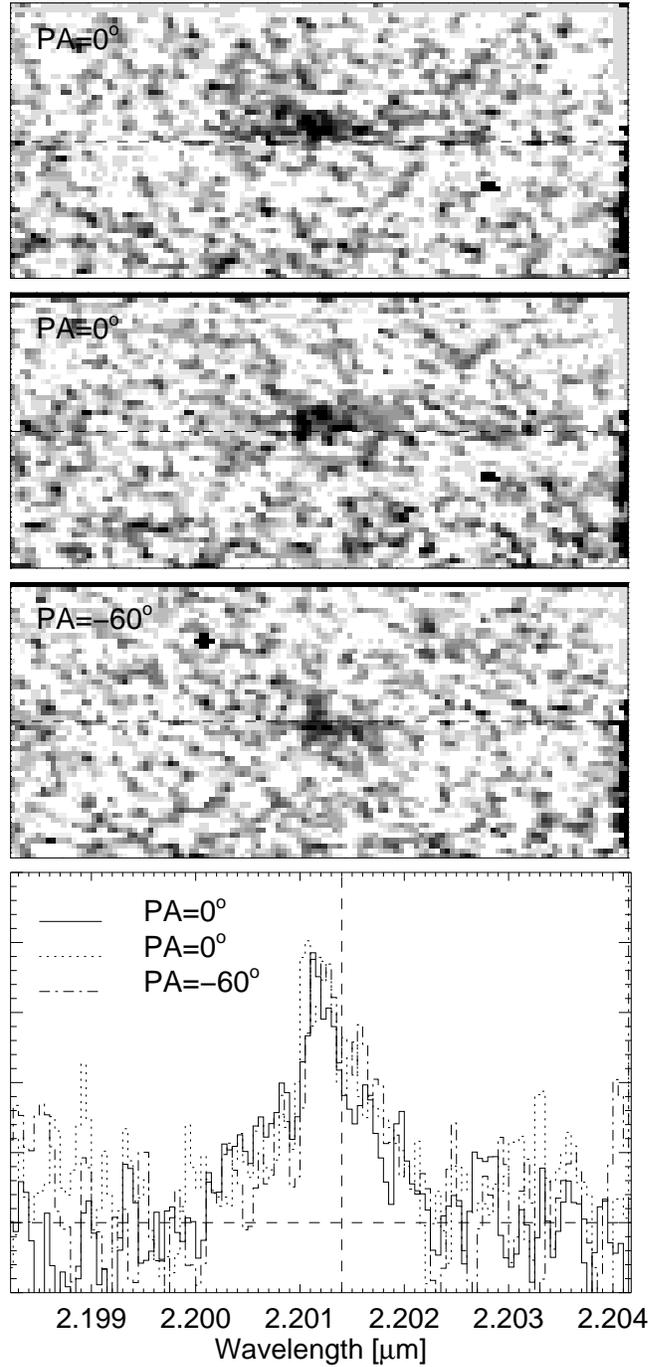}
\caption{The H$\alpha$ emission line from the DLA-galaxy counterpart.  As
in Fig.~\ref{fig:oiii} the upper three figures show the two-dimensional spectra
after SPSF subtraction of the QSO continuum.  In the bottom panel, we plot the
corresponding one-dimensional spectra. The vertical dashed line shows the
predicted position of the H$\alpha$ line from the redshift determined from
low-ionisation absorption lines. The horizontal dashed line indicates the
zero-flux level. The shift in the peak position of the H$\alpha$ line
relative to the position expected from the redshift of low-ionisation absorption 
lines is 25 km $^{-1}$. The origin of this apparently significant shift is 
unclear. The spectra have been rescaled to
have the same peak count-level.
\label{fig:halpha} }
\end{figure}

We clearly detect Ly$\alpha$ emission in the DLA trough in all three
individual spectra. The strongest component of the Ly$\alpha$ emission
in the two $PA=0^\mathrm{o}$ spectra is clearly asymmetric and
redshifted by $\sim$200 km s$^{-1}$ relative to the redshift of
low-ionisation absorption lines (see below). There is also a weak, but
significantly detected, component blueshifted by a similar amount (see
Fig.~\ref{fig:vel} below). The impact parameter in these spectra is
$0.57\pm0.03$ arcsec. From the $PA=-60^\mathrm{o}$ spectrum, which has
been taken under the worst seeing, we measure a redshift which is
consistent with the measurement from the $PA=0^\mathrm{o}$ spectra.
The maximum velocity shift we can expect with a dispersion of 0.25 \AA
\ per pixel and a slit width of 9 pixels is about 2.3 \AA, which at
the wavelength of Ly$\alpha$ would correspond to 170 km s$^{-1}$. The
impact parameter in the $PA=-60^\mathrm{o}$ spectrum is $-0.20\pm0.10$
arcsec. From just two position angles it is not always possible to
uniquely triangulate the position of the galaxy counterpart. However,
in this case we can obtain a fairly secure position (see right panel
of Fig.~\ref{fig:strategy}). The DLA galaxy must be located at an
impact parameter of $\sim$0.8 arcsec and at a position angle of
$\sim40^\mathrm{o}$. As the seeing (as measured around the wavelength
of the redshift Ly$\alpha$ line) degraded substantially from the first
(0.8 arcsec) to the last spectrum (1.6 arcsec) we cannot get an
additional constraint from comparing the counts in the line to the QSO
continuum (this ratio is close to constant in all three spectra).

Integrating over the line profile, we derive a flux of
$f=10.0\times10^{-17}$ erg s$^{-1}$ cm$^{-2}$ in the best of our three
spectra (see Table~\ref{linefluxes}). This constitutes a lower
limit to the total flux due to slit-loss. The corresponding Ly$\alpha$
luminosity is $L=4.3\times10^{42}$ erg s$^{-1}$. Converting this to a
star-formation rate (SFR) using standard case-B recombination theory
and the relation between H$\alpha$ luminosity and SFR from Kennicutt
(1998) gives SFR $= 4$ $M_{\sun}$ yr$^{-1}$.

To search for rest-frame optical emission lines, we carried out
Spectral Point Spread Function (SPSF) subtraction of the QSO light
(see M\o ller 2000). We
do this by locating spectral regions free of sky emission lines on
each side of the expected positions of the rest-frame optical emission
lines. Over these regions, we calculate the average SPSF. We then
subtract the SPSF over the spectral region covering the emission line
by assuming that the QSO spectrum can be fitted by a linear function
across the position of the emission line. The [OIII] $\lambda$4959,
$\lambda$5007, and H$\alpha$ emission lines are located in regions
free of sky lines in the H- and K-bands. After SPSF subtraction, we
clearly detect the [OIII] $\lambda$5007 line in both $PA=0^\mathrm{o}$
spectra and in the $PA=-60^\mathrm{o}$ spectrum (see
Fig.~\ref{fig:oiii}). The redshift of the [OIII] $\lambda$5007 line is
$z_{\rm abs}=2.35406$. This is consistent with the redshift of
low-ionisation absorption lines (see below) within the error, which is
$\lesssim 5$ km s$^{-1}$. The Full Width at Half Maximum (FWHM) of the
[OIII] emission line is 105 km s$^{-1}$, which means that the line is
reasonably well resolved. Correcting for the instrumental resolution,
the intrinsic width of the line is $\sim$80 km s$^{-1}$. The impact
parameters measured from the [OIII] lines are consistent within about 10\%
with what is measured from the Ly$\alpha$ line.

For the flux of the [OIII] $\lambda$5007 line, we measure
$f=3.1\times10^{-17}$ erg s$^{-1}$ cm$^{-2}$ in the best of our three
spectra (see Table~\ref{linefluxes}), which corresponds to a
luminosity of $L=1.3\times10^{42}$ erg s$^{-1}$ similar to the
luminosities of the other two DLAs detected in [OIII] (Weatherley et
al.\ 2005). We also clearly detect the [OIII] $\lambda$4959 emission
line, but as expected at lower significance.

The H$\alpha$ line is detected in the K-band in all three spectra, but
at significantly lower S/N ratio than the [OIII] line. In
Fig.~\ref{fig:halpha}, we show the line in each of our three spectra.
For the H$\alpha$ line, we measure a flux
of $f=2.5\times10^{-17}$ erg s$^{-1}$ cm$^{-2}$, which corresponds to
a luminosity of $L=1.1\times10^{42}$ erg s$^{-1}$ and SFR $= 10$
$M_{\sun}$ yr$^{-1}$ again using the relation between SFR and
H$\alpha$ luminosity from Kennicutt (1998). The ratio between the
[OIII] and H$\alpha$ luminosities is 1.2, in good agreement with the
estimate of Weatherley et al.\ (2005) based on the models of Kewley \&
Dopita (2002). The ratio between the Ly$\alpha$ and H$\alpha$
luminosities is 4.0 whilst a ratio of about 8.7 is expected from
case-B recombination. This suggests that about 55\% of the Ly$\alpha$
photons are either destroyed by dust or that most of the
Ly$\alpha$ emission is emitted in other directions due to 
radiative transfer effects. In general, scattering effects cause a quite
anisotropic escape of Ly$\alpha$; Laursen et al. (2009b) find that 
Ly$\alpha$ emitters on average
will be 2--4 times more luminous when viewed from the brightest direction
compared to when viewed from the least luminous direction.

Unfortunately, one of the components of the [OII] doublet and the
H$\beta$ line are located on top of bright sky emission lines. We
place an upper limit of the [OII] doublet flux by adding to the
observed spectrum a simulated [OII] doublet for which each component
has the same spatial and spectral width as the [OIII] line. We then
scale the flux of the line and infer the lowest flux for which we
would still have detected the line after SPSF subtraction. The upper
limit to the [OII] flux inferred this way is $8\times10^{-17}$ erg
s$^{-1}$ cm$^{-2}$. Using the relation between [OII] luminosity and
SFR from Kennicutt (1998), this limit implies an upper limit of
SFR$<40$ $M_{\sun}$ yr$^{-1}$. Hence, the non-detection is consistent
with the value of SFR $= 10$ $M_{\sun}$ yr$^{-1}$ inferred from
H$\alpha$.

We stress that there are substantial uncertainties in the absolute
fluxes, probably at least 30\%, both due to the possibility of
slit-loss and due to uncertainties in the flux calibration. This error
propagates also to the derived luminosities and SFRs.

\begin{table}
\caption{Emission line fluxes and impact parameters derived from each of the 
the three individual spectra. The line fluxes should be considered lower limits
due to the possibility of slit-loss. The error bars on the line-fluxes are hard
to quantify as systematic errors from SPSF subtraction and background
subtraction contribute significantly. A conservative estimate is 20\%. The 
formal error from propagated photon and read-out noise is in all cases less
than 10\%. The most 
reliable measurements of the impact parameter are from the Ly$\alpha$ line 
as this measurement is unaffected by systematic errors from the SPSF subtraction.
\label{linefluxes}}
\begin{center}
\begin{tabular}{lrrr}
\hline
\hline
Line & PA   & Flux & b \\
     & (degrees)    & (erg $^{-1}$ cm$^{-2}$) & (arcsec) \\
\hline
Ly$\alpha$  &      0  & 10.0$\times$10$^{-17}$ & 0.57 \\ 
Ly$\alpha$  &      0  & 9.0$\times$10$^{-17}$ & 0.57 \\ 
Ly$\alpha$  &  $-60$  & 8.7$\times$10$^{-17}$ & $-0.20$ \\ 
OIII      &      0  & 3.1$\times$10$^{-17}$ & 0.51 \\  
OIII      &      0  & 2.9$\times$10$^{-17}$ & 0.47 \\ 
OIII      &  $-60$  & 3.1$\times$10$^{-17}$ & $-0.24$ \\ 
H$\alpha$   &      0  & 2.5$\times$10$^{-17}$ & 0.56 \\ 
H$\alpha$   &      0  & 2.4$\times$10$^{-17}$ & 0.37 \\ 
H$\alpha$   &  $-60$  & 2.8$\times$10$^{-17}$ & $-0.23$\\ 
\hline
\end{tabular}
\end{center}
\end{table}

\subsection{Absorption properties of the DLA}

The QSO spectrum displays a rich set of absorption lines at $z =
2.354$. We detect transition lines from HI, CII, CII*, CIV, OI, MgI,
MgII, AlII, AlIII, SiII, SiIV, SII, MnII, FeII, NiII and ZnII. CrII
lines are also detected but they are blended with telluric
absorption. By fitting a Voigt profile to the damped HI Ly$\alpha$
line {in the spectrum of the quasar combining all three exposures}, we
determine a neutral hydrogen column density of $\log N_{\rm HI}
=20.65\pm0.05$ (see Fig.~\ref{fig:dla}). This is in good agreement
with the value $\log N_{\rm HI} =20.57\pm0.28$ found in the automatic
analysis of the lower resolution SDSS spectrum by Noterdaeme et al.\
(2009).

In order to estimate the overall metallicity of the DLA, we first use
the SiII $\lambda$1808 transition line as it is weakly saturated and
because silicon is known to deplete little onto dust grains. From the
first moment of the line, we derive a precise redshift of $z_{\rm abs}
= 2.35409$. We also measure a rest-frame line EW of $0.215\pm0.014$
\AA \ which, based on the optically-thin line approximation,
corresponds to a lower limit of [Si/H] $= -0.61\pm0.06$. Here we adopt
the solar photosphere abundances from Asplund et al.\ (2009). Hence,
our selection based on the SiII $\lambda$1526 line indeed has resulted
in a metallicity well above 0.1 solar.

\begin{figure}
\includegraphics[width=0.48\textwidth]{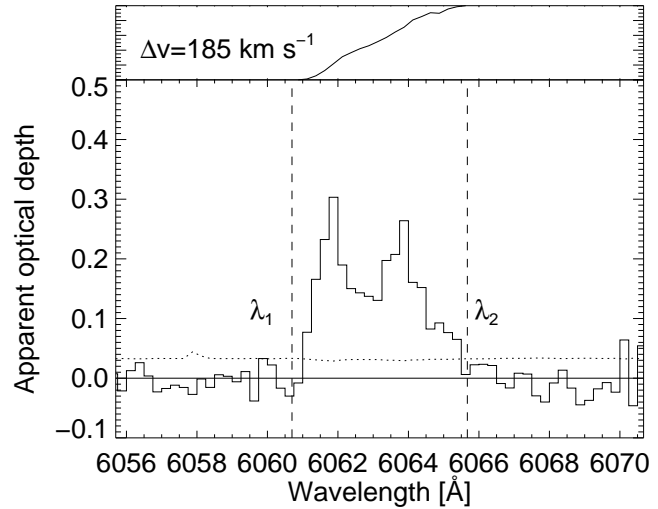}
\caption{The profile of the SiII $\lambda$1808 line. The line profile
consists of two main components separated by about 93 km s$^{-1}$.
The velocity width of the line, measured following the method of
Ledoux et al.\ (2006), is 185 km s$^{-1}$. $\lambda_1$ and
$\lambda_2$ are the start and end wavelengths used to integrate the
profile. \label{fig:profile} }
\end{figure}

\begin{figure}
\psfig{file=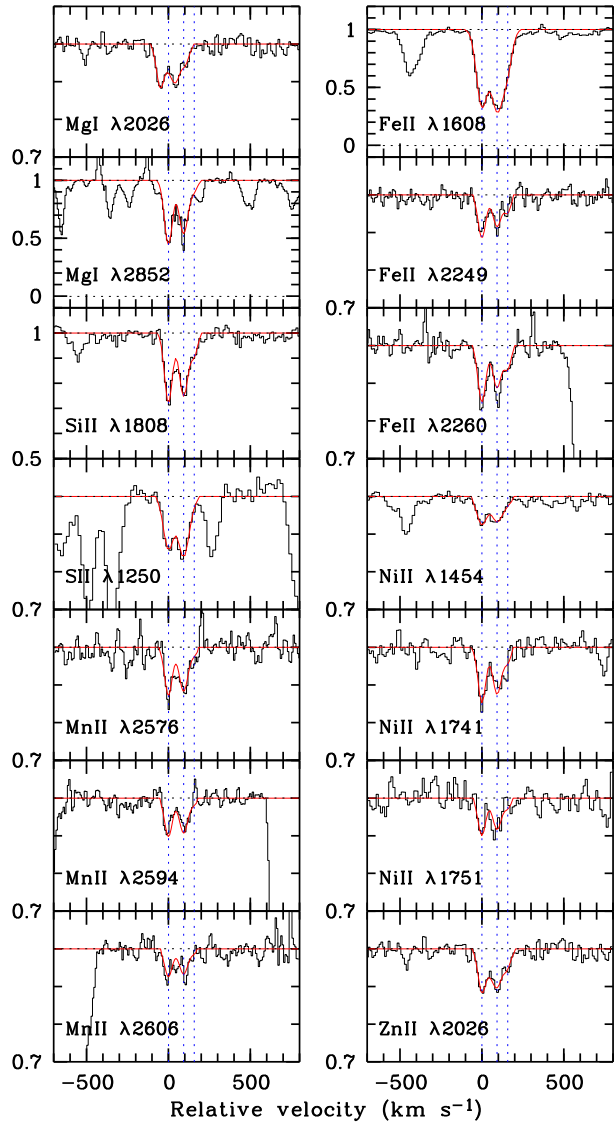,scale=0.69,clip=}
\caption{Results of Voigt-profile fits to low-ionisation lines from
  the $z_{\rm abs}=2.354$ DLA towards Q\,2222$-$0946.
  \label{fig:metallines} }
\end{figure}

We next perform Voigt-profile fitting to measure accurate column
densities for silicon and other elements. As seen from
Fig.~\ref{fig:profile} and Fig.~\ref{fig:metallines}, the metal-line
profiles are well-resolved with widths of $\sim 200$ km s$^{-1}$. The
instrumental resolution is 64 and 45 km s$^{-1}$ for the uvb and vis
arm spectra respectively. This implies that, for a typical S/N of 50,
one can constrain the broadening parameter of individual components
reliably down to a factor of about five smaller ($b\sim 6$ km
s$^{-1}$) than the instrumental broadening. Therefore, as long as the
fitted components have $b$ values larger than about 10 km $s^{-1}$,
there is no problem of hidden saturation. Moreover, as we are
considering weak absorption lines (see Fig.~\ref{fig:metallines}) the
measured column densities depend weakly on $b$ if at all. The
inclusion of one stronger transition line in the fit (FeII
$\lambda$1608) does not change the results because the FeII column
densities are well-constrained by the weak $\lambda$2249 and
$\lambda$2260 lines. Three components are needed to reach Chi$^2$
values of the order of one: two main components separated by about 93
km s$^{-1}$ (see Fig.~\ref{fig:profile}) and one additional, redder
and weaker component (this component is most clearly seen in SiII
$\lambda$1808, ZnII $\lambda$2026 and FeII
$\lambda\lambda$2249,2260). In the following, we refer to the
blueshifted component as 'a' and to the main redshifted component as
'b'. The fits are shown in Fig.~\ref{fig:metallines} and the measured
column densities are given in Table~\ref{metallines}. The
Voigt-profile fit results imply a Si metallicity only slightly higher
than based on the optically-thin line approximation (see above):
[Si/H] $= -0.51\pm0.06$. For Zn which, like Si, is little depleted
onto dust grains, we find [Zn/H] $= -0.46\pm0.07$. For elements which
are more sensitive to dust depletion (e.g., Meyer \& Roth 1990;
Pettini et al.\ 1997; Ledoux et al.\ 2002), we find [Fe/H] $=
-0.99\pm0.06$, [Ni/H] $= -0.85\pm0.06$ and [Mn/H] $= -1.23\pm0.06$,
implying significant dust depletion for these elements.

\begin{table}
\caption {Ionic column densities in individual components of the DLA system at
$z_{\rm abs}=2.354$.\label{metallines}}
\begin{center}
\begin{tabular}{llll}
\hline
\hline
Ion & Transition & $\log N\pm\sigma _{\log N}$ & $b\pm\sigma _b$\\
    & lines used &                            & (km s$^{-1}$)  \\
\hline
\multicolumn{4}{l}{$z_{\rm abs}=2.35341$}  \\
MgI  & 2026,2852        & 12.61$\pm$0.02  &  19.2$\pm$1.2  \\
SiII & 1808             & 15.31$\pm$0.03  &  19.2$\pm$1.2  \\
SII  & 1250             & 14.83$\pm$0.10  &  19.2$\pm$1.2  \\
MnII & 2576,2594,2606   & 12.52$\pm$0.04  &  19.2$\pm$1.2  \\
FeII & 1608,2249,2260   & 14.83$\pm$0.03  &  19.2$\pm$1.2  \\
NiII & 1454,1741,1751   & 13.69$\pm$0.05  &  19.2$\pm$1.2  \\
ZnII & 2026             & 12.43$\pm$0.06  &  19.2$\pm$1.2  \\
\hline
\multicolumn{4}{l}{$z_{\rm abs}=2.35445$}  \\
MgI  & 2026,2852        & 12.49$\pm$0.04  &  26.1$\pm$3.6  \\
SiII & 1808             & 15.30$\pm$0.04  &  26.1$\pm$3.6  \\
SII  & 1250             & 14.92$\pm$0.09  &  26.1$\pm$3.6  \\
MnII & 2576,2594,2606   & 12.53$\pm$0.04  &  26.1$\pm$3.6  \\
FeII & 1608,2249,2260   & 14.74$\pm$0.04  &  26.1$\pm$3.6  \\
NiII & 1454,1741,1751   & 13.66$\pm$0.06  &  26.1$\pm$3.6  \\
ZnII & 2026             & 12.38$\pm$0.07  &  26.1$\pm$3.6  \\
\hline
\multicolumn{4}{l}{$z_{\rm abs}=2.35517$}  \\
MgI  & 2026,2852        & 11.53$\pm$0.20  &   9.3$\pm$1.3  \\
SiII & 1808             & 14.68$\pm$0.12  &   9.3$\pm$1.3  \\
SII  & 1250             & $<13.45^{\rm a}$&   9.3$\pm$1.3  \\
MnII & 2576,2594,2606   & 11.59$\pm$0.28  &   9.3$\pm$1.3  \\
FeII & 1608,2249,2260   & 14.34$\pm$0.10  &   9.3$\pm$1.3  \\
NiII & 1454,1741,1751   & 13.01$\pm$0.20  &   9.3$\pm$1.3  \\
ZnII & 2026             & 11.71$\pm$0.29  &   9.3$\pm$1.3  \\
\hline
\end{tabular}
\end{center}
$^{\rm a}$ 3$\sigma$ upper limit.
\end{table}

To compare the kinematics of the absorption line profiles with that of
other DLAs, we follow the procedure of Ledoux et al.\ (2006) and
calculate the line-profile velocity width, $\Delta v$, as
$c\left[\lambda(95\%)-\lambda(5\%)\right]/\lambda_0$, where
$\lambda(5\%)$ and $\lambda(95\%)$ are the wavelengths corresponding
to, respectively, the 5 and 95 percentiles of the apparent optical
depth distribution, and $\lambda_0$ is the first moment (the average)
of this distribution (see fig.~1 of Ledoux et al.\ 2006). We again
choose SiII $\lambda$1808 as it is a low-ionisation transition line
and the line is weakly saturated. The apparent line optical depth and
the derived velocity width are shown in Fig.~\ref{fig:profile}. We
infer a velocity width of 185 km s$^{-1}$ in good agreement with the
velocity-metallicity relation for DLAs (Ledoux et al.\ 2006).

\section{Discussion}

\subsection{The DLA-galaxy counterpart}

Simultaneous detection of the Ly$\alpha$, [OIII] and H$\alpha$
emission lines have not been seen for DLA galaxies before and it is
rarely seen for other types of high-$z$ galaxies. The galaxy
counterpart of the DLA appears to be a vigorously star-forming galaxy
with a SFR of 10 $M_{\sun}$ yr$^{-1}$ (this is a lower limit due to
the possibility of slit-loss and dust absorption). 
Compared to the H$\alpha$ luminosity
function for $z\approx2$ galaxies determined by Hayes et al.\ (2010),
the galaxy has an H$\alpha$ luminosity of $0.1L^*_\mathrm{H\alpha}$.
From the impact parameter of the object of 0.8
arcsec, we can infer a length scale of at least 6 kpc for the extent
of the neutral hydrogen associated with the galaxy. The metallicity of
the gas is relatively high, 2/5 solar. This is a higher metallicity
than any of the measured metallicities for DLAs at similar redshifts
in the large compilation presented in Prochaska et al.\ (2003). It is
consistent with the model described in Fynbo et al.\ (2008) to have a
high metallicity despite a relative large impact parameter as the
metallicity gradient in this model is shallow for the brightest
galaxies (see also Boissier \& Prantzos 2001). Unfortunately, the
constraint on the metallicity based on the strong, rest-frame optical
emission lines is too loose. Based on the measured [OIII] luminosity,
the upper limit on the [OII] luminosity, and assuming a flux ratio of
2.88 between H$\alpha$ and H$\beta$ (the canonical case B ratio assuming
no dust absorption), we infer a limit of $\log{R_{23}}
< 1.15$, which is fulfilled at all metallicities (see Kewley \& Dopita
2002 for details).

The spatial profiles of both the Ly$\alpha$ and the [OIII] line are
consistent with the SPSF so the star-forming region of the galaxy is
much more compact than the extent of the neutral gas.

\begin{figure}
\includegraphics[width=0.42\textwidth]{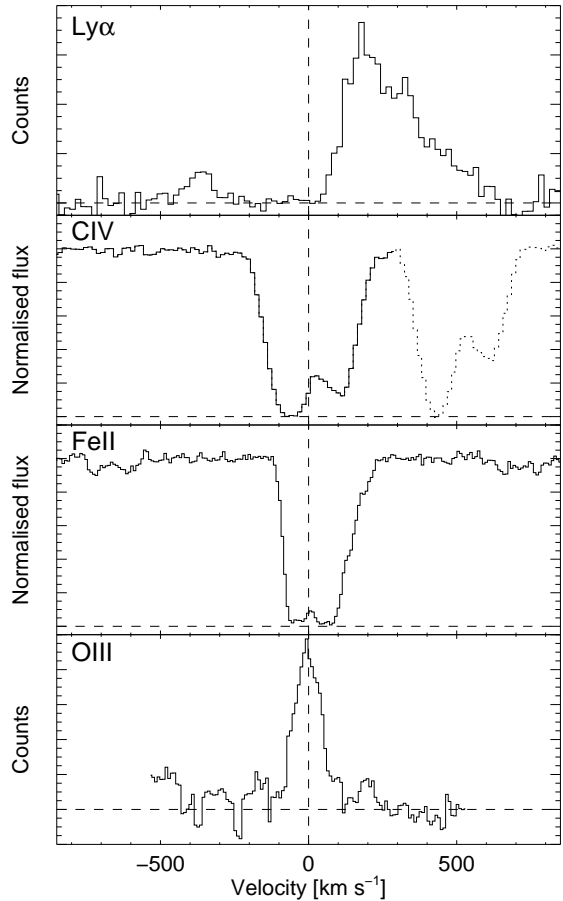}
\caption{The velocity profiles of the [OIII] emission line (bottom),
  low-ionisation absorption lines represented by the FeII
  $\lambda$2600 line (lower middle), high-ionisation absorption lines
  represented by the CIV $\lambda$1548 line (upper middle), and the
  Ly$\alpha$ emission line (top). The Ly$\alpha$ and [OIII] emission
  lines have been extracted after SPSF subtraction of the QSO
  continuum. The zero-point of the velocity scale is relative to a
  systemic redshift of $2.35406$ measured from the [OIII] emission
  line. \label{fig:vel} }
\end{figure}

The line luminosities, impact parameter and metallicity of the
DLA-galaxy counterpart are similar to what has been found for the few
other $z\gtrsim2$ DLAs detected in emission (M\o ller \& Warren 1998;
M\o ller et al. 2002, 2004). The rest-frame EW of the MgII
$\lambda$2796 absorption line is 2.7 \AA, and hence the system
fulfills the selection criterion of the strong $z\sim1$ MgII absorbers
studied by Bouch{\'e} et al.\ (2007, see also M\'enard et al.\ 2009).
In that study, galaxy counterparts are detected in H$\alpha$ emission
for 67\% of the studied systems. In terms of SFR, the galaxy
counterpart of the DLA studied here falls within the range (1--20
$M\sun$ yr$^{-1}$) found for $z\sim1$ MgII absorbers. In terms of its
emission-line luminosities and MgII EW, the DLA-galaxy counterpart is
also similar to the $z=0.4$--0.7 systems studied by Noterdaeme et al.\
(2010).

\subsection{Kinematics}

This DLA system is unique for having accurate determinations of the
Ly$\alpha$ and [OIII] emission-line redshifts and of low- and
high-ionisation absorption-line redshifts from the interstellar medium
(ISM) of the absorbing galaxy. This makes it possible to infer
kinematical information that is rarely available in such detail (see
Fig.~\ref{fig:vel}). The [OIII] emission line is thought to trace the
systemic redshift of the galaxy as [OIII] photons originate from the
HII regions in the galaxy and are unaffected by resonant scattering
effects. The width of the [OIII] line, corrected for the instrumental
resolution, is 80 km s$^{-1}$. This width must be related to the
dynamical mass of the star-forming component of the galaxy, but as we
do not have a size we cannot infer a mass. The line width is
comparable to what has been found for bright LBGs at slightly larger
redshifts (Pettini et al.\ 2001; see also Weatherley \& Warren 2003,
2005).

The low-ionisation absorption lines are represented by the saturated
FeII $\lambda2600$ line in Fig.~\ref{fig:vel}. The absorption profile
is characterised by two main components separated by about 93 km
s$^{-1}$. The total width of the absorption line profile is 185 km
s$^{-1}$, i.e. substantially wider than the [OIII] emission line. The
difference may possibly be interpreted as the effect of a galactic
wind, but the two velocity widths also probe quite different scales in
the galaxy (a sightline through the neutral gas vs. the star-forming
component of the galaxy). The absorption redshift is very similar to
the systemic redshift, which is remarkable given the impact parameter
of $\sim$6 kpc. We note that, due to the offset within the slit
between the QSO and the DLA galaxy (see Fig.~\ref{fig:strategy}), we
expect a velocity shift. Assuming a roughly quarter slit width offset,
we expect a shift of 20 km s$^{-1}$, so this is a small effect. In the
few previous DLAs for which this information is available, blueshifts
of several 100 km s$^{-1}$ have been found between systemic and
low-ionisation absorption lines (Weatherley et al.\ 2005). For LBGs,
low-ionisation ISM absorption lines are also typically blueshifted
(Pettini et al.\ 2001). The small velocity shift observed in the
present case may be an indication that we observe a disk-like system
with a line-of-sight that is close to parallel with the rotation axis.
The profile of the high-ionisation CIV line follows that of the
low-ionisation lines, although component a is stronger than component
b which is opposite to the relative strength of the two components for
the low-ionisation lines.

\subsection{The Ly$\alpha$ emission-line profile}

The Ly$\alpha$ line is a resonant line, and the physical processes
governing its formation are quite different from the metal lines. Due
to the effects of scattering, Ly$\alpha$ photons have a complex path
out of galaxies \citep[][and references therein]{lau07}. The profile
of the line shown in the top panel of Fig.~\ref{fig:vel} is clearly
asymmetric and remarkably the blue edge of the profile is close to the
position of Ly$\alpha$ at the systemic redshift. Even more striking is
the presence of a small peak on the blue side of the systemic
redshift. This can be understood as a result of resonant scattering.
In general, due to the high opacity of neutral hydrogen for a photon
in the line centre, Ly$\alpha$ photons have to diffuse in frequency to
either the blue or the red side. Thus, to first order the line should
be characterised by two symmetric peaks with the central minimum
indicating the redshift (Harrington 1973; Urbaniak \& Wolfe 1981). Although this has
been observed on several occasions \citep[e.g.][]{ven05,tap07},
typically observations show that the blue peak is missing, leaving
only the asymmetric red part with the distinct steep rise in intensity
close to the line centre and more shallow decrease farthest from the
centre. 

Various scenarios may explain the weakness of the blue peak. In
particular, if large outflows of gas are present, as may be expected
in the presence of a starburst, in the reference frame of the
outflowing hydrogen atoms blue photons are shifted towards the line
centre, while red photons are shifted away from resonance. Thus, the
photons will escape more easily after having drifted to the red side.
While no analytical solution exists for this scenario\footnote{Except
  in the academic case of $T = 0$ \citep{loe99}.}, it has been
investigated numerically, both in the case of a homologous expansion
\citep{dij06} and that of a thin, circumgalactic shell of gas
\citep{ver06}. Even if outflows are not present, after the photons
escape the galaxy, the redshift due to the Hubble expansion of the
intergalactic medium may remove part of the blue peak. At $z \simeq
2.4$, the average transmission bluewards of the Ly$\alpha$ line is
80--85\% \citep{son04}. However, since in the vicinity of the galaxy
the expansion is slower due to the gravitational potential of the galaxy,
more absorption may occur close to the line centre. In the case of
infalling gas, even the red peak of the spectrum may be affected
\citep{dij07}.

Considering the relatively high metallicity inferred for the system and the
apparent depletion of refractory elements, the presence of dust is anticipated.
Due to the path length of the Ly$\alpha$ radiation being increased by resonant
scattering, dust may suppress the Ly$\alpha$ line more than other lines. Even
for ``grey'' dust, the line is not affected uniformly, the wings being
influenced more than the centre \citep{lau09b}. The reason is that the wings
are comprised of photons originating in the very dense regions that have to
diffuse far from the line centre in order to escape, while photons near the
line centre must come from low-density regions. Since the dense regions are
where also most of the stars are, and hence most of the dust, these photons
have a higher probability of being absorbed.

If we wish to understand the physical conditions responsible for the
formation of the observed Ly$\alpha$ line, we must take all of these
effects into account. An exact fit to the Ly$\alpha$ emission-line
profile requires careful numerical modeling, varying a multitude of
physical, and often degenerate, parameters. \citet{ver08} successfully
modeled a number of Ly$\alpha$ profiles by varying the temperature,
column density, expansion velocity and dust contents of a thin shell.
Analytical solutions exist only for homogeneous, isothermal, and
static configurations of gas, with either a central or evenly
distributed source of light. In the case of a homogeneous gas, each of
the two peaks are fairly symmetric about their maxima, but taking into
account the full range of densities, and the correlation of Ly$\alpha$
emission with these densities, often results in significantly more
skewed peaks. This is the result of different parts of the spectrum
originating in physically distinct regions. A realistic scenario of
galactic outflows probably lies somewhere in between that of the
homologously expanding sphere and that of a thin shell. From Fig.~8 of
\citet{lau09a}, one sees that in the case of the expanding sphere with
a maximum velocity $V$ of 20 km s$^{-1}$, the red peak maximum is
approximately twice as high as the blue peak maximum, while for $V =
200$ km s$^{-1}$ the blue peak is missing completely. In the case of a
shell, from Fig.~14 of \citet{ver06} slightly larger expansion
velocities can occur while still allowing the blue peak to be seen.

Considering the above discussion on the radiative transfer of
Ly$\alpha$ radiation, we find that the observed emission is consistent
with originating at the same redshift as the systemic redshift, having
scattered its way out of an inhomogeneous ISM of gas temperature $T
\sim 10^4$ K, neutral hydrogen column density comparable to that
inferred from the DLA absorption, as well as a fair amount of dust
absorbing $\sim$1/2 of the photons, expanding at a typical velocity of
several tens to $\sim$100 km s$^{-1}$, and further having reduced its
blue side somewhat by the surrounding IGM.

\section{Conclusions: The relation between emission- and absorption-selected galaxies}

The progress on understanding the relation between emission- and
absorption-selected galaxies at high redshift has been slow. Today,
more than 20 years after the onset of systematic studies of DLAs
(Wolfe et al.\ 1986; Smith et al.\ 1989) and 15 years after the
discovery of LBGs we still have very little empirical basis for
establishing the relation between the two. Here we have presented the
results from the first DLA observed in an X-shooter survey targeting
metal-rich DLAs, namely the $z_{\rm abs}=2.354$ DLA towards
Q\,2222$-$0946. We confirm that our pre-selection of metal-rich DLAs
based on the strength of the SiII $\lambda$1526 line works: the
inferred metallicity of the system is [Zn/H] = $-0.46\pm0.07$. We also
clearly detect the galaxy counterpart of the DLA absorber in
Ly$\alpha$, [OIII] and H$\alpha$ emission. The DLA-galaxy counterpart
is a vigorously star-forming galaxy with a SFR of at least 10
$M_{\sun}$ yr$^{-1}$. The line-of-sight probes the neutral gas
associated with this galaxy at an impact parameter of about 6 kpc. The
kinematical information inferred from emission and absorption suggests
that we observe the galaxy with a line-of-sight that is close to
parallel with its rotation axis. From the profile of Ly$\alpha$, we
can infer that there most likely is a galactic wind causing the
blueshifted part of the line to be absorbed.

The implication of this is that we now have both the strategy
(selection of metal-rich DLAs) and the tool (the X-shooter
spectrograph) that will allow us to attack the question of the nature
of DLA-galaxy counterparts.

It would also be interesting to further explore the inverse
experiment, namely to characterise the HI absorption due to LBGs close
to the sightlines to background QSOs (or other LBGs). Adelberger et
al.\ (2005) address this issue, but mainly focusing on metal-line
absorption and at impact parameters of several hundred kpc. In a few
cases, candidate LBGs have been found within a few arcsec from the
background QSOs (Steidel \& Hamilton 1992; Steidel et al.\ 1995), but
only for one of these has confirming spectroscopy been published
(Djorgovski et al.\ 1996; Weatherley et al.\ 2005). In the spectrum of
the background QSO, this galaxy is detected as a sub-DLA.

We need a larger sample before firm conclusions can be drawn, but the
detection of the galaxy counterpart of the DLA towards Q\,2222$-$0946
with properties such as impact parameter and SFR within the
predictions does give us hope that the basic picture presented in the
works of Fynbo et al.\ (2008) and in the hydrodynamical simulations of
Pontzen et al.\ (2008) are correct. We are hence within reach of
firmly establishing the connection between high-$z$ galaxies detected
in emission and absorption.

\section*{Acknowledgments}
The Dark cosmology centre is funded by the DNRF. We thank Christina
Th\"one, Max Pettini, Regina Jorgenson and Andrew Pontzen and our
anonymous referee for helpful comments.


\begin{thebibliography}{99}
\bibitem[Adelberger et al.(2005)]{Adelberger05}
Adelberger, K., Shapley, A. C., Steidel, C. C., et al. 2005, ApJ, 629, 636
\bibitem[Adelberger et al.(2006)]{Adelberger06}
Adelberger, K., Steidel, C. C., Kollmeier, J. A., \& Reddy, N. A. 2006, ApJ, 637, 74
\bibitem[Adelman-McCarthy et al.(2009)]{Adelman09} Adelman-McCarthy, J. K., Ag\"ueros, M. A., Allam, S. S. et al. 2009, ApJS, 175, 297
\bibitem[Asplund et al.(2009)]{Asplund09}
Asplund, M., Grevesse, N., Sauval, A. J., \& Scott, P. 2009, ARA\&A, 47, 481
\bibitem[Atek et al.(2009)]{Atek09}
Atek, H., Kunth, D., Schaerer, D., et al. 2009, A\&A, 506, L1
\bibitem[Bohlin \& Gilliland(2004)]{Bohlin04}
Bohlin, R. C. \& Gilliland, R. L. 2004, AJ, 128, 3053
\bibitem[Bohlin(2007)]{Bohlin07}
Bohlin, R. C. 2007, ASPC, 364, 315
\bibitem[Boissier \& Prantzos(2001)]{BP01}
Boissier, S. \& Prantzos, N. 2001, MNRAS, 325, 321
\bibitem[Bouche et al.(2007)]{Bouche07}
Bouch{\'e}, N., Murphy, M. T., P{\'e}roux, C., et al. 2007, ApJ, 669, L5
\bibitem[Bouwens et al.(2009)]{Bouwnes09}
Bouwens, R. J., Illingworth, G. D., Oesch, P. A., et al. 2009, ApJL, in press (arXiv:0909.1803B)
\bibitem[Cen \& Ostriker(1999)]{Cen09}
Cen, R. \& Ostriker, J. P. 1999, ApJL, 519, L109
\bibitem[Charlot \& Fall(1991)]{CharlotFall}
Charlot, S. \& Fall, S. M. 1991, ApJ, 378, 471
\bibitem[Christensen et al.(2007)]{Christensen07}
Christensen, L., Wisotzki, L., Roth, M. M., et al. 2007, A\&A, 468, 487
\bibitem[Colbert \& Malkan(2002)] {colbert02}
Colbert, J. W. \& Malkan, M. A. 2002, ApJ, 566, 51
\bibitem[Dijkstra et al.(2006)] {dij06} Dijkstra, M., Haiman, Z., \& Spaans, M. 2006, ApJ, 649, 14
\bibitem[Dijkstra et al.(2007)] {dij07}
Dijkstra, M., Lidz, A., \& Wyithe, J. S. B. 2007, MNRAS, 377, 1175
\bibitem[Djorgovski et al.(1996)]{Djorgovski96}
Djorgovski, S. G., Pahre, M. A., Bechtold, J., \& Elston, R. 1996, Nature, 382, 234
\bibitem[Fynbo et al.(1999)]{Fynbo99}
Fynbo, J. P. U., M\o ller, P., \& Warren, S. J. 1999, MNRAS, 305, 849
\bibitem[Fynbo et al.(2008)]{Fynbo08}
Fynbo, J. P. U., Prochaska, J. X., Sommer-Larsen, J., Dessauges-Zavadsky, M. \& M\o ller, P. 2008, ApJ, 683, 321
\bibitem[Goldoni et al.(2006)]{Goldoni06}
Goldoni, P., Royer, F., Fran{\c c}ois, P., et al. 2006, SPIE, 6269, 80
\bibitem[Haehnlet et al.(2000)]{Haehnlet00}
Haehnelt, M. et al. 2000, ApJ, 534, 594
\bibitem[Harrington(1973)]{Harrington1973}
Harrington, J. P. 1973, MNRAS, 162, 43
\bibitem[Hayes et al.(2010)]{Hayes2010}
Hayes, M., Schaerer, D., \& \"Ostlin, G. 2010, A\&A, 509, L5
\bibitem[Heinm\"uller et al.(2006)]{Heinmuller06}
Heinm\"uller, J., Petitjean, P., Ledoux, C., Caucci, S. \& Srianand, R. 2006, A\&A, 449, 33
\bibitem[Kaplan et al.(2010)]{Kaplan2010}Kaplan, K. F., Prochaska, J. X., 
Herbert-Fort, S., Elisson, S. L., \& Dessauges-Zavadsky, M. 2010, PASP, 122, 619
\bibitem[Kelson(2003)]{Kelson03}Kelson, D. D. 2003, PASP, 115, 688
\bibitem[Kennicutt(1998)]{Kennicutt98}
Kennicutt, R. C. 1998, ARA\&A, 36, 189
\bibitem[Kewley \& Dopita(2002)]{Kewley02}
Kewley, L. J. \& Dopita, M. 2002, ApJS, 142, 35
\bibitem[Kulkarni et al.(2006)]{Kulkarni06}
Kulkarni, V. P., Woodgate, B. E., York, D. G., et al. 2006, ApJ, 636, 30
\bibitem[Laursen \& Sommer-Larsen(2007)]{lau07}
Laursen, P. \& Sommer-Larsen, J. 2007, ApJL, 657, L69
\bibitem[Laursen et al.(2009a)]{lau09a} Laursen, P., Razoumov, A. O., \&
Sommer-Larsen, J. 2009a, ApJ, 696, 853 
\bibitem[Laursen et al.(2009b)]{lau09b}
Laursen, P., Sommer-Larsen, J., \& Andersen, A. C. 2009b, ApJ, 704, 1640
\bibitem[Ledoux et al.(2002)]{Ledoux02}
Ledoux, C., Bergeron, J., \& Petitjean, P. 2002, A\&A, 385, 802
\bibitem[Ledoux et al.(2006)]{Ledoux06}
Ledoux, C., Petitjean, P., Fynbo, J. P. U., M\o ller, P., \& Srianand, R., 2006, A\&A, 457, 71
\bibitem[Leibundgut \& Robertson(1998)]{iznogood98}
Leibundgut, B., \& Robertson, J. G. 1999, MNRAS, 303, 711
\bibitem[Loeb \& Rybicki(1999)]{loe99}
Loeb, R. \& Rybicki, G.~B. 1999, ApJ, 524, 527
\bibitem[Lowenthal et al.(1995)]{Lowenthal95}
Lowenthal, J., Hogan, C. J., Green, R. F., et al. 1995, ApJ, 451, 484
\bibitem[M\'enard et al(2009)]{Menard09}
M\'enard, B., Wild, V., Nestor, D., Quider, A., \& Zibetti, S. 2009, MNRAS, submitted (arxiv:0912.3263)
\bibitem[Meyer \& Roth(1990)]{MR90}
Meyer, D. M. \& Roth, K. C. 1990, ApJ, 363, 57
\bibitem[M\o ller (2000)]{Moller00}
M{\o}ller, P. 2000, {\it The Messenger}, 99, 31
\bibitem[M\o ller \& Warren(1993)]{Moller93}
M{\o}ller, P. \& Warren, S. J. 1993, A\&A, 270, 43
\bibitem[M\o ller \& Warren(1998)]{Moller98}
M{\o}ller, P. \& Warren, S. J. 1998, MNRAS, 299, 661
\bibitem[M\o ller et al.(2002)]{Moller02}
M\o ller, P., Warren, S. J., Fall, Fynbo, J. P. U, \& Jakobsen, P. 2002, ApJ, 574, 51
\bibitem[M\o ller et al.(2004)]{Moller04}
M\o ller, P., Fynbo, J. P. U., \& Fall, S. M. 2004, A\&AL, 422, L33
\bibitem[Noterdaeme et al.(2009)]{Noterdaeme09}
Noterdaeme, P., Petitjean, P., Ledoux, C., \& Srianand, R. 2009, A\&A, 505, 1087
\bibitem[Noterdaeme et al.(2010)]{Noterdaeme10}
Noterdaeme, P., Srianand, R., \& Mohan, V. 2010, MNRAS, 403, 906
\bibitem[Pei \& Fall(1995)]{Pei95}
Pei, Y. C., \& Fall, S. M. 1995, ApJ, 454, 69
\bibitem[Pettini et al.(1997)]{Pettini97}
Pettini, M., King, D. L., Wmith, L. J., \& Hunstead, R. W. 1997, ApJ, 478, 536
\bibitem[Pettini et al.(2001)]{Pettini01}
Pettini, M., Shapley, A. E., Steidel, C. C., et al. 2001, ApJ, 554, 981
\bibitem[Pettini(2006)]{Pettini06}
Pettini, M. 2006, In: Proceedings of the Vth Marseille International Cosmology
conference (Edited by V. LeBrun, A. Mazure, S. Arnouts and D. Burgarella), p. 319
\bibitem[Pontzen et al.(2008)]{Pontzen08}
Pontzen, A., Governato, F., Pettini, M., et al. 2008, MNRAS, 390, 1349
\bibitem[Pontzen \& Pettini(2009)]{Pontzen09}
Pontzen, A, \& Pettini, M. 2009, MNRAS, 393, 557
\bibitem[Prochaska et al.(2003)]{X03}
Prochaska, J. X., Gawiser, E., Wolfe, A.M., Castro, S., \& Djorgovski, S. G. 2003, ApJL, 595, L9
\bibitem[Prochaska et al.(2005)]{X05}
Prochaska, J. X., Herbert-Fort, S., \& Wolfe, A. M. 2005, ApJ, 635, 123
\bibitem[Prochaska et al.(2008)]{X08}
Prochaska, J. X., Chen, H.-W., Wolfe, A., et al. 2008, ApJ, 672, 59
\bibitem[Schaye et al.(2001)]{Schaye01}
Schaye, J. 2001, ApJL, 559, L1
\bibitem[Smith et al.(1989)]{Smith89}
Smith, H. E., Cohen, R. D., Burns, J. E., Moore, D. J., Uchida, B. A. 1989, ApJ, 347, 87
\bibitem[Sommer-Larsen \& Fynbo(2008)]{Jesper08}
Sommer-Larsen, J., \& Fynbo, J. P. U. 2008, MNRAS, 385, 3
\bibitem[Songaila(2004)]{son04}
Songaila, A. 2004, ApJ, 127, 2598
\bibitem[Steidel \& Hamilton(1992)]{Steidel92} 
Steidel, C. C., \& Hamilton, D. 1992, AJ, 104, 941 
\bibitem[Steidel et al.(1995)]{Steidel95} 
Steidel, C. C., Pettini, M., \& Hamilton, D. 1995, AJ, 110, 2519
\bibitem[Steidel et al.(1996)]{Steidel96} Steidel, C. C., 
Giavalisco, M., Pettini, M., Dickinson, M., \& Adelberger, K. L., 1996, ApJL, 462, L17
\bibitem[Steidel et al.(2004)]{Steidel04} Steidel, C. C., Shapley, A., Pettini, M., et al. 2004, ApJ, 604, 534
\bibitem[Tapken et al.(2007)]{tap07}
Tapken, C, Appenzeller, I., Noll, S., et al. 2007, A\&A, 467, 63
\bibitem[Tescari et al.(2009)]{Tescari09} 
Tescari, E., Viel, M., Tornatore, L., \& Borgani, S. 2009, MNRAS, 397, 411
\bibitem[Urbaniak \& Wolfe(1981)]{Urbaniak81} 
Urbaniak, J. J \& Wolfe, A. M. 1981, ApJ, 244, 406
\bibitem[van Dokkum(2001)] {Dokkum01} van Dokkum, P. G. 2001, PASP, 113, 1420
\bibitem[Venemans et al.(2005)] {ven05} Venemans, B., R\"ottgering, H. J. A., Miley, G. K., et al. 2005, A\&A, 431, 793
\bibitem[Verhamme et al.(2006)] {ver06} Verhamme, A., Schaerer, D., \& Maselli, A. 2006, A\&A, 460, 397
\bibitem[Verhamme et al.(2008)]{ver08}
Verhamme, A., Schaerer, D., Atek, H., \& Tapken, C. 2008, A\&A, 491, 89
\bibitem[Warren \& M\o ller(1996)]{Warren96}
Warren, S. J. \& M\o ller, P. 1996, A\&A, 311, 25
\bibitem[Weatherley \& Warren(2003)]{Weatherley03}
Weatherley, S. J. \& Warren, S. J. 2005, MNRAS, 345, L29
\bibitem[Weatherley \& Warren(2005)]{Weatherley05b}
Weatherley, S. J. \& Warren, S. J. 2005, MNRAS, 363, L6
\bibitem[Weatherley et al.(2005)]{Weatherley05}
Weatherley, S. J., Warren, S. J., M\o ller, P, et al. 2005, MNRAS, 358, 985
\bibitem[Wolfe et al.(1986)]{Wolfe86} Wolfe, A. M., Turnshek, D. A., Smith, H. E., \& Cohen, R. D. 1986, ApJS, 61, 249
\bibitem[Wolfe et al.(2005)]{Wolfe05} Wolfe, A. M., Gawiser, E., \& Prochaska, J. X. 2005, ARA\&A, 43, 861
\bibitem[Zwaan et al.(2005)]{Zwaan05} Zwaan, M., van der Hulst, J. M., Briggs, F. H., Verheijen, M. A. W., \& Ryan-Weber, E. 2005, MNRAS, 364, 1467 \bibitem[Zwaan et al.(2008)]{Zwaan08}
Zwaan, M., Walter, F., Ryan-Weber, E., et al. 2008, AJ, 136, 2886
\end{thebibliography}
\end{document}